\documentclass[%
 aip,
 amsmath,amssymb,
 reprint,%
]{revtex4-1}

\usepackage{graphicx}
\usepackage{dcolumn}
\usepackage{bm}

\usepackage[utf8]{inputenc}
\usepackage[T1]{fontenc}
\usepackage{mathptmx}

\begin{document}

\preprint{AIP/123-QED}

\title[Complexity plane]{Exploiting the impact of ordering patterns in the Fisher-Shannon complexity plane
}

\author{David Spichak}
 \email{david.spichak@gmail.com}
\author{Andr\'es Aragoneses}%
\email{aaragoneses@ewu.edu}
\affiliation{ 
Department of Physics, Eastern Washington University, Cheney, WA 99004, USA 
}%

\date{\today}

\begin{abstract}
The Fisher-Shannon complexity plane is a powerful tool that represents complex dynamics in a two-dimensional plane. It locates a dynamical system based upon its entropy and its Fisher Information Measure (FIM). It has been recently shown that, by using ordinal patterns to compute permutation entropy and FIM, this plane unveils inner details of the structure underlying the complex and chaotic dynamics of a system. In order to compute FIM, the order of the patterns is very relevant. We analyze in detail the impact of the sorting protocol used to calculate FIM using ordinal patterns. We show the importance of a suitable choice, which can lead to saving computational resources, but also to unveil details of the dynamics not accessible to other sorting protocols. Our results agree with previous research, and common characteristic fingerprints are found for the different chaotic maps studied. Our analysis also reveals the fractal behavior of the chaotic maps studied. We extract some underlying symmetries that allow us to simulate the behavior on the complexity plane for a wide range of the control parameters in the chaotic regimes.
 
\end{abstract}

\maketitle

\begin{quotation}

Describing and characterizing the complex dynamics of chaotic systems is a challenging task. Some techniques are computationally demanding, and some hidden features of the dynamics are hard to be unveiled. Entropy and Fisher Information Measure (FIM) are two quantifiers that extract global and local information of the dynamics, respectively. The Fisher-Shannon plane combines them to locate the dynamics of a system on a 2D projection. Recent research showed that using ordinal patters to compute the entropy (as permutation entropy) and FIM extracts more detail of the dynamics than otherwise, but the way FIM is defined is sensitive to the order in which we classify the ordinal patterns. In this paper we analyze in detail the impact of the sorting protocol on the structure unveiled by the Fisher-Shannon plane and show preferred sorting protocols in order to extract more detailed information from iterative chaotic maps. We find that all chaotic maps under study present the same fingerprint on the plane. We also show specific orders that help display the fractal behaviour of maps that present period-doubling route to chaos. Finally, we find hidden symmetries that allow us to simulate the generic features of the chaotic maps on the Fisher-Shannon plane.
\end{quotation}

\section{\label{sec:level1}Introduction}

There are several tools that can be used to quantify and characterize chaos, that help distinguish stochasticity from deterministic chaos, and allow to discriminate different dynamical regimes in a complex dynamical system~\cite{2015_Chaos_Kantz,2017_PRL_Politi,2020_Nature_Esposito,2003_RSI_Tracy,2013_SciRep_Aragoneses}. Among them, Entropy and Fisher Information Measure (FIM) are two quantifiers that extract global and local information, respectively, from a complex system. A convenient way to present these two quantifiers is through the Fisher-Shannon complexity plane~\cite{2003_Vignat}, that projects the FIM of the system versus its entropy, localizing the dynamics of the system on a two-dimensional plane. This technique has shown to be able to characterize nonlinear dynamics, and to distinguish between stochastic noise and deterministic chaos~\cite{2012_Phys_LettA_Olivares, 2012_PhysicaA_Olivares, 2014_PLOS_Carpi} by comparing and tracking the locations of dynamical systems on the plane.

In a recent work~\cite{2021_Spichak} we studied the Fisher-Shannon complexity plane of different chaotic maps under the lens of Bandt and Pompe's~\cite{2002_PRL_BP} ordinal patterns approach. We found that, using this ordinal patterns approach to calculate the Fisher-Shannon plane is more powerful than a PDF-based Fisher-Shannon plane, when it comes to unveil the hidden structure of complex dynamical systems. An interesting feature we identified is that most of the non-invertible iterative maps share a common fingerprint on the plane as the control parameter is scanned. We also showed how this technique allows to detect transitions in the dynamical behavior of the chaotic system.

Something to be aware of, though, when computing the location of a complex dynamical system on the Fisher-Shannon plane, is that, while entropy is invariant to how we order the patterns, FIM is sensitive to the chosen order~\cite{2012_PhysicaA_Olivares, 2021_Spichak}. This can lead one system to be projected in different locations on the plane when using different arrangements of the patterns. In this paper we explore the impact of the sorting protocol of the ordinal patterns in extracting information through the Fisher-Shannon complexity plane.

\section{\label{sec:level1}The Fisher-Shannon plane using ordinal patterns.}

\subsection{\label{sec:level2}Permutation Entropy}

In order to compute the entropy and FIM of the time series of a complex dynamical system, we first transform the time series of the dynamics into a sequence of patterns, also known as {\it{words}}. These words are built by comparing consecutive values of the time series~\cite{2002_PRL_BP}. The number of possible words will depend on their dimension, i.e., on how many consecutive values we consider to construct the words. Words are assigned depending on the relative magnitudes of consecutive points in the time series. For example, for dimension $D=2$ we compare two consecutive values, $\{ x_i, x_{i+1}\}$. If $x_i < x_{i+1}$ we assign the word $01$. For $x_{i+1} < x_{i}$ we assign the word $10$. Similarly, for dimension $D=3$, if  $x_i < x_{i+1} < x_{i+2}$ then we assign the word $012$; if  $x_i < x_{i+2} < x_{i+1}$ then we assign the word $021$, and so on. We have a total of $D!$ different words for dimension 
$D$.

From the sequences of patterns we calculate the probability of each word, and
we compute the Permutation Entropy (PE) as
\begin{equation}
    PE=\frac{1}{ln(D!)}\sum_{i=1}^{D!} p_i~ln(p_i)~,
\end{equation}

\noindent where $D$ is the dimension of the words, $D!$ is the number of different possible words, $p_i$ is the probability of the i-th word, and PE is normalized so that $0 \leq PE \leq 1$. Permutation entropy is a measure of the global behavior of the dynamics, robust to changes in the distribution on small scales, and invariant to the way we order the $i=D!$ words. It quantifies the distance of the system to an uncorrelated stochastic process.

This protocol implies a compression of information: words lose the information of the exact values in the time series, but they are able to extract signatures of temporal correlations in the dynamics. This method of calculating the entropy is robust to local perturbations and to experimental noise. It is useful in unveiling long temporal correlations in the dynamics, in finding temporal scales, in distinguishing random behavior from determinism, and in statistically forecasting events~\cite{2014_OptEx_Toomey,2017_Entropy_Bandt,2017_Entropy_Bandt, 2016_PRL_Aragoneses, 2019_Entropy_Bandt, 2020_FrontPhys_Zanin,2018_SciRep_Colet}.

\subsection{\label{sec:level2}Fisher Information Measure}

Fisher Information Measure (FIM) is another quantifier of complexity. It measures the gradient of the distribution. This makes it be sensitive to small, localized changes. For a  distribution with $N$ possible values it can be defined as
\begin{equation}
    FIM = F_0 \sum_{i=1}^{N-1}\left((p_{i+1})^{1/2}-(p_i)^{1/2}\right)^2~,
\end{equation}
\noindent where $F_0 = 1$ if $p_{i^*}=1$ for $i^*=1$ or $i^*=N$, and $p_i^*=0~\forall i \neq i^*$. $F_0=\frac{1}{2}$ otherwise. $\{p_i,~i=1,..,N\}$ is the discrete probability distribution set. FIM is a powerful tool to identify and characterize complexity in nonlinear dynamical systems~\cite{1925_Fisher,1999_PLA_Plastino,2004_Frieden,2018_PRA_Carpi}.

PE and FIM complement each other, as the former extracts information of the dynamics at a global scale, while the latter does it at a local scale. For this reason the Fisher-Shannon plane is a good tool to characterize complex dynamics, distinguish stochasticity from chaos, and differentiate dynamical regimes.

However, the way FIM is defined entails that the order of the probability distribution set affects the computed FIM. For a given set of $N$ different values with probabilities $\{p_i,~i=1,..,N\}$ there are $N!$ possible options of sorting them. Each one of the possible sorting protocols can provide a different value of FIM. Olivares et al.~\cite{2012_PhysicaA_Olivares} studied the two most common ordering criteria, Lehmer and Keller, for the logistic map, using words of dimension $D=6$. For the windows of regularity of the map, they found that both FIM coincide, but in the chaotic regions each sorting criteria gives a different FIM. They detected a linear relationship between both criteria for the period doubling regions present in the dynamics, but no clear relationship for more chaotic regions.

For words of dimension $D=3$ there are $D!=6$ possible words, and $(D!)!=720$ possible orders for these words. This makes a detailed analysis of all possible orders an impossible challenge for dimensions $D \geq 4$. Here we present results of a detailed analysis of the impact on the Fisher-Shannon plane of the ordering criteria for words for dimension $D=3$.

\section{\label{sec:level1}The Fisher-Shannon plane for different sorting criteria}

By using the ordinal patterns approach, one can reveal different levels of complexity and structure when using words of different dimensions. In the case of the Fisher-Shannon plane, higher dimensions can extract more details from the dynamics. Nevertheless, words of dimension $D=5$ get the same qualitative details as words of higher dimension (see Fig. 2 from Ref.~\cite{2021_Spichak}).

\begin{figure}[ht]
\centerline{
\includegraphics[width=7.5cm]{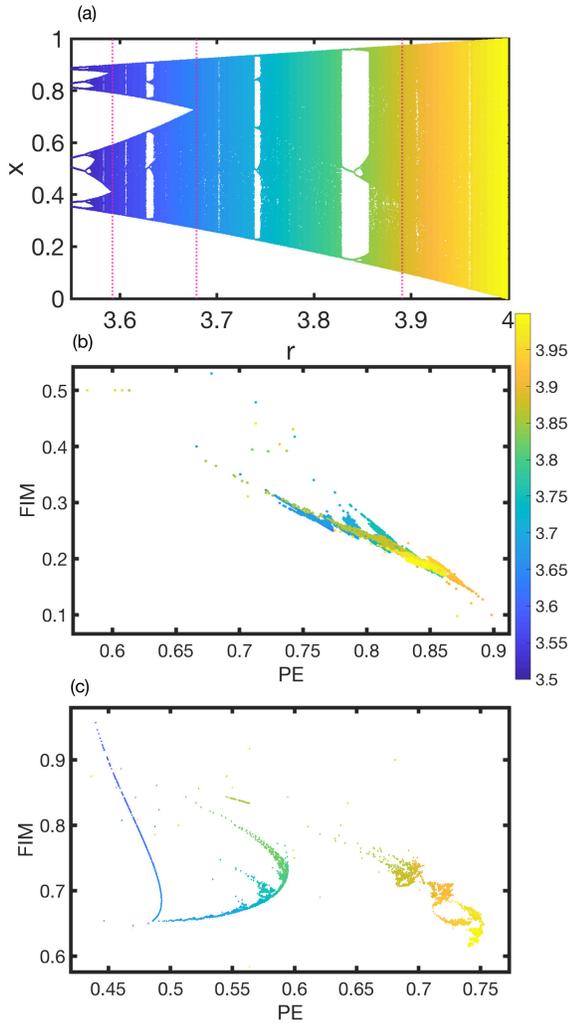}
}
\caption{(a) Bifurcation diagram of the logistic map. Dashed lines indicate where symmetry breaking takes place (explained in the main text). (b) Fisher-Shannon plane for the logistic map using ordinal patterns of dimension D=3. (c) Fisher-Shannon plane for the logistic map using ordinal patterns of dimension D=4. FIM has been computed using Lehmer sorting protocol. Control parameter range is $3.5 \leq r \leq 4.0$. We  use a color scale to identify $r$ in the figures.}
\label{bifurcation}
\end{figure}

Figure \ref{bifurcation} presents the bifurcation diagram of the logistic map, $x_{n+1}=rx_n(1-x_n)$, and the Fisher-Shannon plane for words of dimension $D=3$ and $D=4$. It is clear how the higher dimension plot presents a more detailed structure, distinguishing regions based upon their dynamics. For example, while for dimension $D=3$ most of the $r$-values lie on the same region showing low structure, following a straight-line shape on the plane, with blue dots (lower $r$ values) and yellow dots (higher $r$ values) occupying the same area; for dimension $D=4$ the complexity plane distinguishes different regions as the system deploys a clear structured pattern on the plane as we scan the control parameter $r$.

\begin{figure}[ht]
\centerline{
\includegraphics[width=9cm]{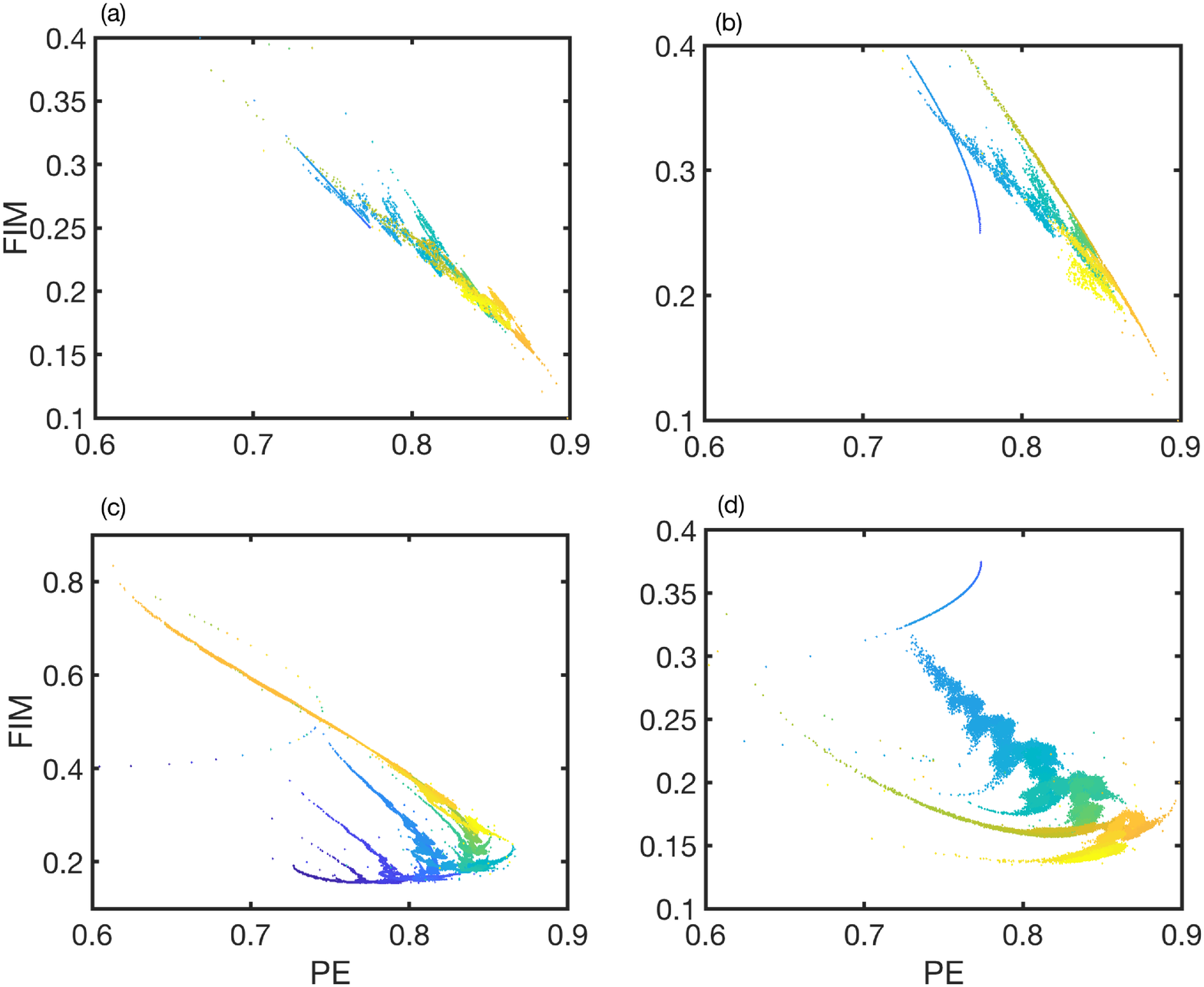}
}
\caption{Fisher-Shannon plane for the logistic map using words of dimension $D=3$. Four different ordering protocols are presented: (a) Lehmer (SA-123456), (b) Keller (SA-125346), (c) SA$^{\dagger}$ (SA-431652), and (d) SA$^*$ (SA-241536). See main text for a description of SA labels.}
\label{logistic}
\end{figure}

It makes sense that higher dimensionality picks up more details of the dynamics, as we are considering more possible words, each one of them extracting a particular feature of the temporal correlations in the time series. But reordering the words before we compute FIM has also an impact on the level of detail we can extract from the plane. Figure~\ref{logistic} presents the Fisher-Shannon plane for the logistic map using words of dimension $D=3$ for four different ways of sorting the words before computing PE and FIM. In it, we can appreciate how different the structure portrayed is. While there is no change in the values of PE across the sorting protocols, FIM does change, and it affects the vertical axis of the plots.
 
Figure~\ref{logistic}a corresponds to Lehmer sorting protocol, \ref{logistic}b to Keller, \ref{logistic}c to SA$^{\dagger}$, and \ref{logistic}d to SA$^*$ (the labeling of these sorting protocols to be described later). While the FIM range in Figs.~\ref{logistic}a, \ref{logistic}b, and \ref{logistic}d are comparable, Fig.~\ref{logistic}c needs a wider vertical axes to accommodate the dispersion in FIM values for SA$^{\dagger}$, indicating the suitability of this sorting protocol to extract dynamical structure in the system.

In order to differentiate regions that have different dynamical features it would be beneficial to find that sorting protocol that spreads out the chaotic map the most on the plane. This would be optimum to locate those values of the control parameter for which the dynamical behaviour is similar, in nearby regions of the geography of the plane, separated from those values of the control parameter for which the dynamical behaviour is more different to them.

There are 720 ways of sorting the six words of dimension $D=3$. In order to label them we follow the next criteria: we assign a number to each one of the six words as
\begin{center}
\begin{tabular}{c|c}
Ordinal pattern & Word label \\
\hline
012 & 1  \\
021 & 2  \\
102 & 3 \\
120 & 4 \\
201 & 5  \\
210 & 6  \\
\end{tabular}
\end{center}

Then we assign the sorting array (SA) corresponding to the order of the label of the words to each specific sorting protocol. After this, Lehmer protocol is SA-123456; Keller protocol is SA-125346, and so on:

\begin{center}
\begin{tabular}{c|c|c|c}
 Lehmer & Keller & SA$^*$ & SA$^{\dagger}$\\
 SA-123456 & SA-125346 & SA-241536 & SA-431652\\
\hline
012 & 012 & 021 & 120\\
021 & 021 & 120 & 102\\
102 & 201 & 012 & 012\\
120 & 102 & 201 & 210\\
201 & 120 & 102 & 201\\
210 & 210 & 210 & 021\\
\end{tabular}
\end{center}

 We find that the protocols SA-241536 ($SA^*$) and SA-431652 ($SA^{\dagger}$) are specially convenient to see detailed structure (see Figs.~\ref{logistic}c and \ref{logistic}d). This sorting arrays scatter FIM values on the plane differentiating the internal features more than most sorting arrays.
 
Of all the 720 different sorting arrays there are some for which their projection on the Fisher-Shannon plane coincide. A display of all the sorting arrays for the logistic map can be found in Ref.~\cite{logistic_video, logistic_cusp_video}.

\begin{figure}[ht]
\centerline{
\includegraphics[width=7.5cm]{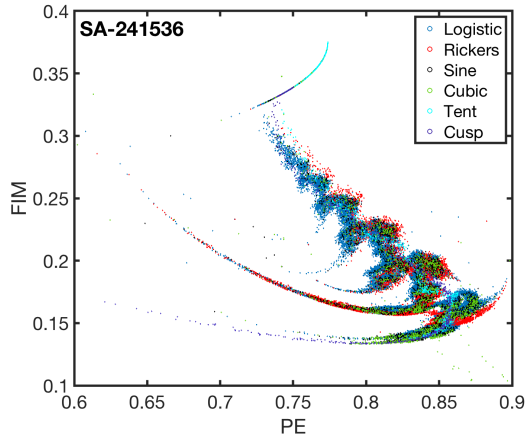}
}
\caption{Fisher-Shannon plane for words of dimension $D=3$ using SA-241536 ($SA^*$) for the logistic, Ricker's, sine, cubic, tent, and cusp maps. Just as happened for SA-Lehmer~\cite{2021_Spichak}, they all share the same fingerprint on the plane.}
\label{noninvertible}
\end{figure}

In a previous work~\cite{2021_Spichak}, using SA-Lehmer, we found that most non-invertible maps leave the same fingerprint on the Fisher-Shannon plane, even when this fingerprint changes with the dimensionality of the words. This commonality is also present for the other sorting arrays. Figure \ref{noninvertible} shows the Fisher-Shannon plane for the logistic, Ricker's, sine, cubic, tent, and cusp maps, for $SA^*$ (SA-241536). In it, all maps cover the same region and they feature the same structure, pointing at akin dynamical complexity. As it was found for Lehmer, the tent and cusp maps present a more simple structure that only covers the skeleton of the more complex structure of the other maps. This behaviour is present for other sorting arrays (not shown).

\section{\label{sec:level1}Complex structure in the Logistic map}

As can be seen from Fig.~\ref{logistic}a, sorting the words using SA-Lehmer does not extract as much structure as other protocols, such as $SA^*$ for example (see Fig.~\ref{logistic}d), where the plot for the logistic map covers a wider region on the plane, some recurrent behavior can also be appreciated, and the most chaotic values (yellow) do not overlap the less chaotic ones (blue and green), as happens for SA-Lehmer. There is also a well defined curve for lower $r$ values (blue in the figure), that is not obvious with SA-Lehmer. This range of $r$ values goes from the onset of chaos ($r=3.569$) to the merging of the two branches in the bifurcation diagram ($r=3.679$). Even though the system is chaotic for these $r$ values, this well defined curve is indicating the presence of some internal symmetry in the dynamics.

\begin{figure}[ht]
\centerline{
\includegraphics[width=8cm]{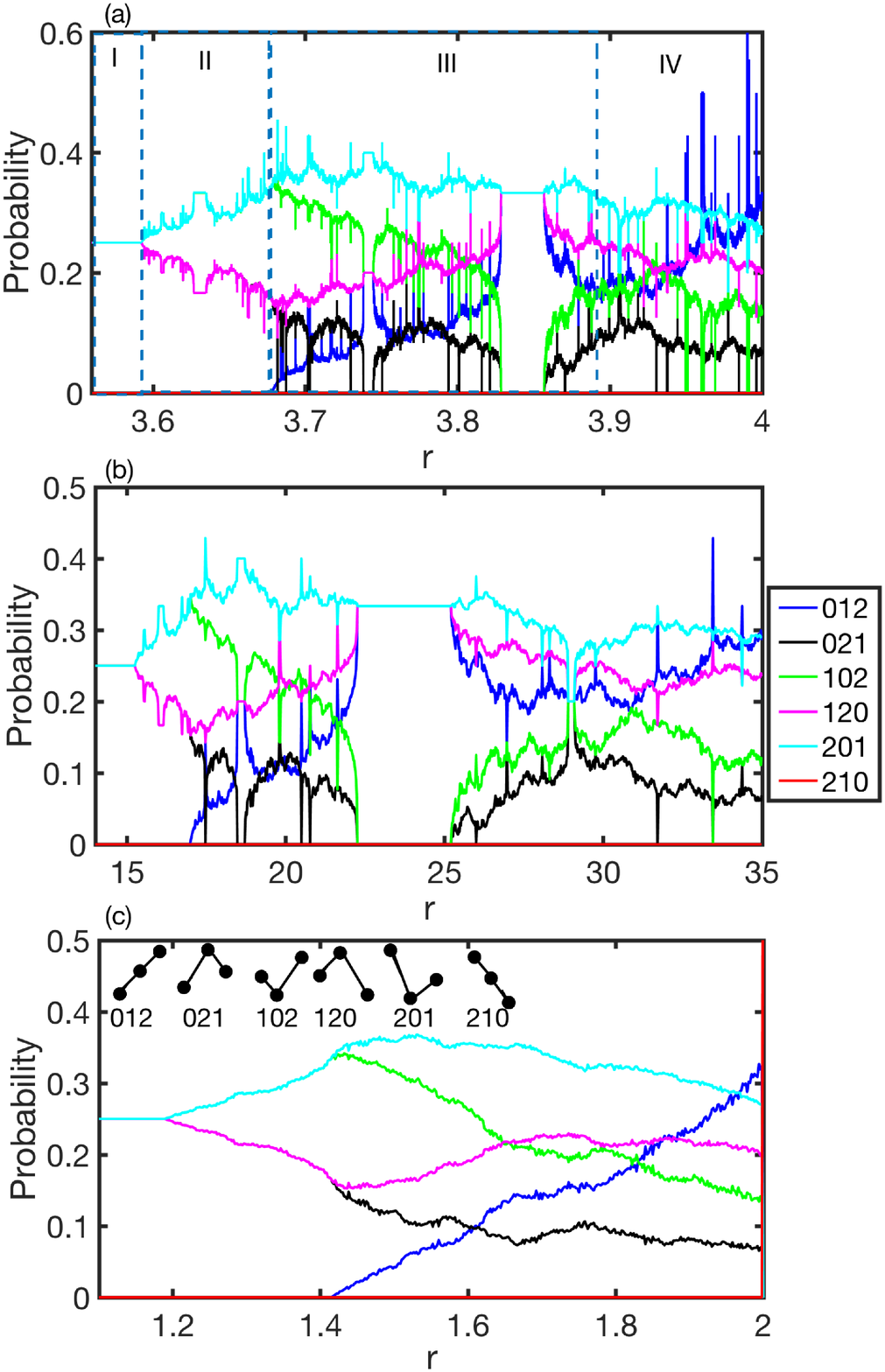}
}
\caption{Words probabilities ($D=3$) versus control parameter $r$, for the logistic (a), Ricker's (b), and tent maps (c). In the logistic map (a) three regions are distinguished indicating different internal symmetries in the complex dynamics. In (c) visual representations of the words are depicted.}
\label{words}
\end{figure}

In order to explore the details of the symmetries on the dynamics in this regime, we plot the probabilities of the six words of dimension $D=3$ versus the control parameter of each map. Figure~\ref{words} shows those probabilities for the logistic, Ricker's, and tent maps. All three figures share common features as the control parameter is scanned. This general behaviour is also present in other non-invertible maps (not shown).

From Fig.~\ref{words} we see that the system starts with high symmetry, portrayed in the words probabilities for all these maps  with one cluster of four words. Words 021, 102, 120, and 201 have the same probability, while words 012 and 210 are forbidden (210  is forbidden for the whole range). After that, this high symmetry breaks into a lower symmetry, the cluster splits into two smaller clusters: 021-120 and 102-201. These two clusters then break down so that each word from previous clusters has a different probability, and the word 012 starts to have non-zero probability.

The fact that words are grouped in clusters indicates the presence of determinism and internal symmetries in the dynamics. In each of those regions we can find a correlation among probabilities more restrictive than the default one:
\begin{equation}
P_j=1-\sum_{i\neq j}P_i~.
\label{eq_words}
\end{equation}

There are four differentiated chaotic regions in these maps. They are indicated in Fig.~\ref{words}a for the case of the logistic map:
\begin{itemize}
    \item
Region I ($r<3.592$) contains one cluster and two forbidden words ($P_1=P_6=0$, $P_2=P_3=P_4=P_5=\frac{1}{4}$). The system on the complexity plane is fully determined in this range, with a defined value for PE and for FIM.

Words in this region are grouped such that each word in the cluster is the time inverse of another word in the same cluster ($102\rightarrow 201$, $012\rightarrow 210$, ...), but also the time and intensity inverse ($102\rightarrow 021$, $012\rightarrow 012$, ...). This indicates that the dynamics presents mirror symmetry (temporal reversibility) and rotational symmetry~\cite{2021_Gunther}.

The term rotational symmetry introduced in Ref.~\cite{2021_Gunther} comes from the fact that, by rotating 180 degrees one of the words, as seen in Fig.~\ref{words}c, we obtain the other word in the cluster.

\item Region II ($3.592<r<3.679$) contains two clusters and two forbidden words
\begin{equation}
\begin{array}{lc}
P_1=P_6=0~,\\
P_2=P_4\neq P_3=P_5~.
\end{array}
\label{eq_constraint_II}
\end{equation}

The system here has one degree of freedom, as we only need to know one of the probabilities to determine all the remaining ones.

In this region the dynamics presents only mirror (temporal) symmetry, as each cluster is formed by those words that are the time inverse of the each (012-210; 102-201; 021-120).

\item Region III ($3.679<r<3.891$) has no clusters and only one forbidden word ($P_6=0$), but the probabilities of the five remaining words are constrained by the following correlations:
\begin{equation}
\begin{array}{lc}
P_2=\frac{1-3P_1}{2}-P_3~,\\
P_4=P_1+P_2~,\\
P_5=P_{1}+P_{3}~.
\end{array}
\label{eq_constraint_III}
\end{equation}
The system here has two degrees of freedom. We only need to determine the probabilities of two words (for example $P_1$ and $P_3$) to determine uniquely its position on the complexity plane.

\item In region IV ($3.891<r \leq 4.0$) we have not found constraints in the probabilities, other than that of Eq.~\ref{eq_words} and $P_6=0$. The probabilities are not correlated.
\end{itemize}

\begin{figure}[ht]
\centerline{
\includegraphics[width=6.5cm]{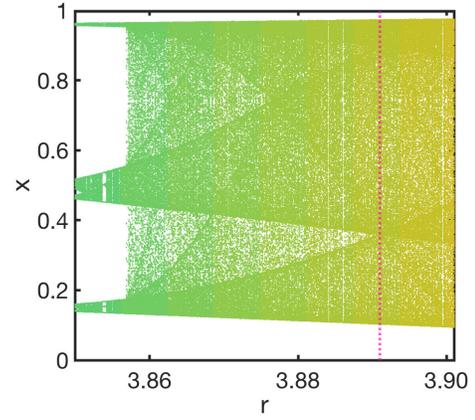}
}
\caption{Bifurcation diagram for the Logistic map around the third symmetry breaking, $r=3.891$, indicated by the dotted line.}
\label{symmetry}
\end{figure}

The same differentiation of regions is found for the other non-invertible maps under consideration. The dotted lines in Fig.~\ref{bifurcation}a indicate the r-values where there is a symmetry breaking. Two of them ($r=3.592$ and $r=3.679$) can be related to the merging of two wide regions in the bifurcation map. The third one is not obvious. If we zoom in the $r=3.891$ region (see Fig.~\ref{symmetry}) we can see that there is another merging-like behavior at this value, although the bifurcation diagram is much more complex here, and the dynamics much less restrictive.

Because of these correlations found in the different regions, we can simulate the Fisher-Shannon complexity plane without using the iterative equation of any specific map. All the maps that contain these correlations between words probabilities should lie on the same region, they should present the same fingerprint on the plane.

\begin{figure}[ht]
\centerline{
\includegraphics[width=9cm]{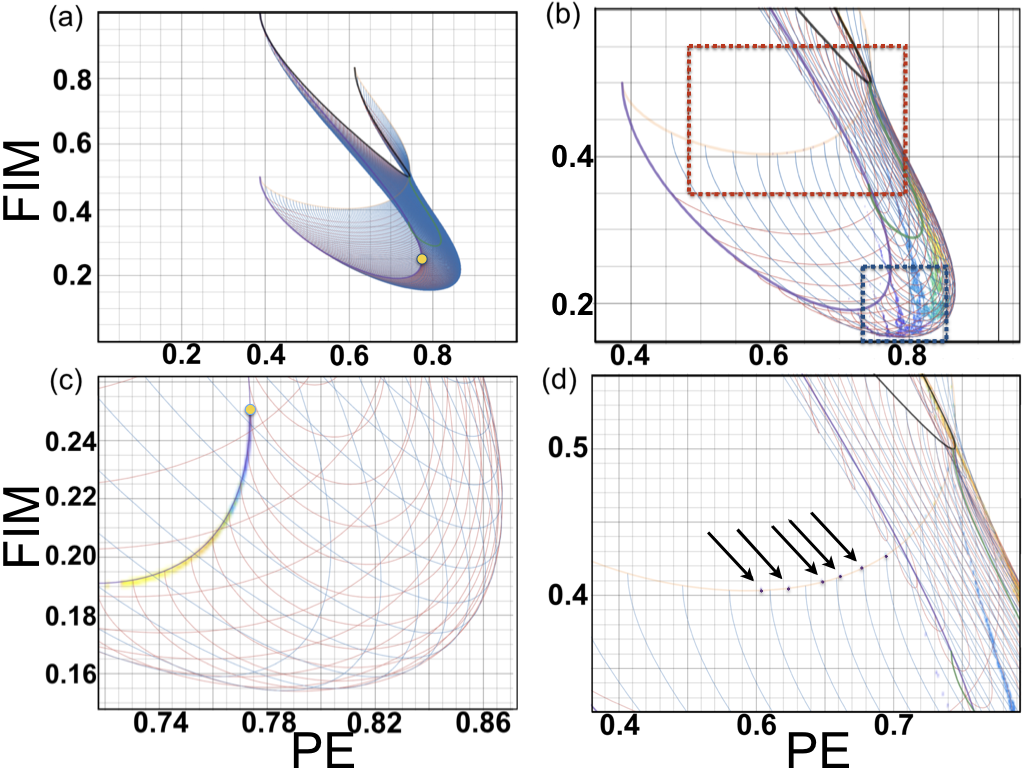}
}
\caption{{\bf{(a)}} Numerical simulation of the complexity plane using the underlying constraints (Eqs.~\ref{eq_constraint_II} and \ref{eq_constraint_III}), for SA$^{\dagger}$ and $D=3$. {\bf{(b)}} Logistic map is shown overlapped with the numerical simulation. {\bf{(c)}} zooms in the blue dashed rectangle in (b). Region II for the logistic map is also shown overlapped with the numerical simulation.
{\bf{(d)}} zooms in the red dashed rectangle in (b). Arrows indicate the dots corresponding to the windows of periodicity in the logistic map in regions II and III, which lie on top of one of the boundaries of the simulation.}
\label{numerical}
\end{figure}

For region I, because the probabilities are fixed by $P_2=P_3=P_4=P_5=\frac{1}{4}$, there is only one option for PE and for FIM (for SA$^{\dagger}$, PE=0.7737 and FIM=0.25). This means that all this region is described by a single dot on the plane (yellow dot in Figs.~\ref{numerical}a, \ref{numerical}c). This dot will be different for different sorting arrays (SA), but it will coincide for some.

For region II we have a richer behavior, that allows for several combinations to satisfy that  $P_2=P_4 \neq P_3=P_5$. For this region the system leaves the dot on the FIM-PE plane and starts to explore other values on the plane.
If we scan $0 \leq P_2=P_4 \leq \frac{1}{2}$ and $P_3=P_5=\frac{1}{2}-P_2$, we obtain all allowed projections on the complexity plane. Plotting this constraint on the plane for SA$^{\dagger}$ we obtain the well defined curved line starting at the point defining region I shown in Fig.~\ref{numerical}c. This curved line defines the allowed values for region II of the different chaotic maps. Of course, while scanning the probabilities from lowest to highest in the simulation, we obtain a continuous line. In the case of the chaotic maps, the dots that define that line do not go uniformly from end to end of the line, but they are scattered on top of the line as the control parameter of each map is scanned.

Looking at Fig.~\ref{noninvertible} we can appreciate that all the maps occupy parts of this curved line in the top part of the plane, but none of them occupies the whole allowed curved line, and not all the maps occupy the exact same points on the curved line, as the probabilities explored by each map are different, despite preserving the constraint among words (Eq.~\ref{eq_constraint_II}).

Because of the high symmetry imposed by the constraints in this region, there are only eight different groups of sorting arrays, which display eight different sections of the curved line~\cite{logistic_video}.

It is worth noting that our numerical simulation, based on the probabilities, defines the allowed values on the plane, but it does not indicate which values of that line will be actually represented by each chaotic map. That is given by the combinations of probabilities that each map is covering, which is not the whole range. Our numerical representation is the maximum set of values on the plane that the chaotic map can take, it is a bound to the possible region where the map can be found on the plane.

For region III we impose the less restrictive constraint from Eq.~\ref{eq_constraint_III}, and plot all the possible values (see Fig.~\ref{numerical}a). Because of the more relaxed constraint, the possible landscape for this region is wider. Nevertheless, it can be appreciated how the projection of the logistic map is wrapped inside the allowed region of the simulation (Fig.~\ref{numerical}b).

The logistic map, in its chaotic regime presents windows of periodicity. These windows correspond to more deterministic and symmetric dynamics which decrease PE and increase FIM. Those windows of periodicity can be spotted on the plane as dots (see Fig~\ref{numerical}d). Those dots lie at one of the boundaries of the allowed region defined by the simulation.

\section{\label{sec:level1}Fractal behaviour unveiled by the Words-Fisher-Shannon plane}

Sorting arrays SA$^*$ and SA$^{\dagger}$ (505 and 307 in Ref.~\cite{logistic_video}) are two, out of the 720, that present more detail on the Fisher-Shannon plane than most of the others (see Figs.~\ref{logistic}c, \ref{logistic}d). They both spread out the landscape of the maps on the plane, they do not present overlap of different regions, and they show different visual structure for different regions of the control parameter. By inspecting any of them (see Fig.~\ref{noninvertible} for SA$^*$) one can appreciate certain repetition of the features that is not revealed by SA-Lehmer (See Fig.~\ref{bifurcation}b). This repeated patterns suggests that this method can also unveil the self-similarity present in the the logistic map.

\begin{figure}[ht]
\centerline{
\includegraphics[width=7cm]{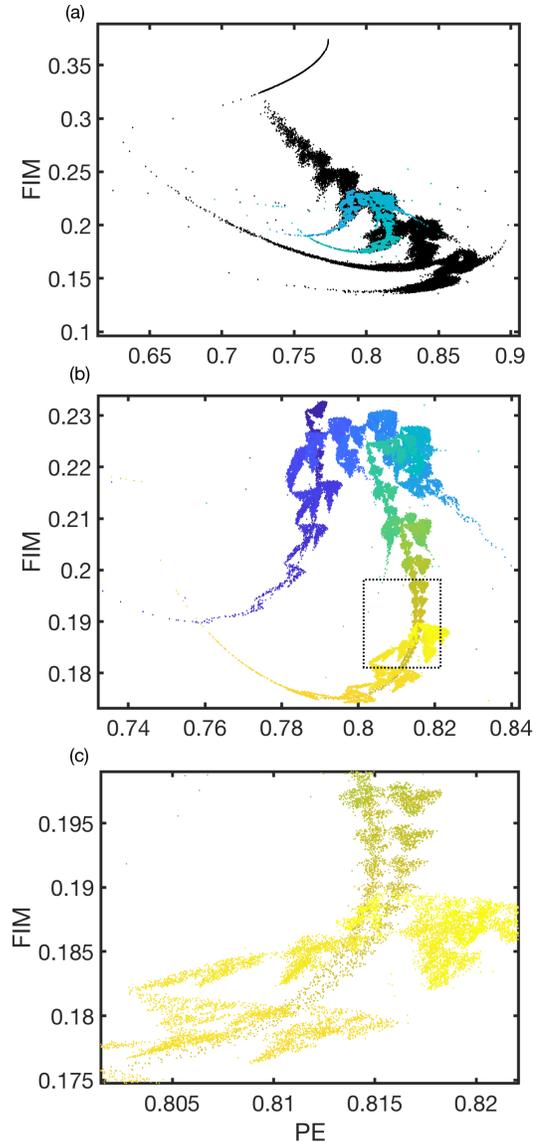}
}
\caption{Fisher-Shannon plane using SA$^*$ for the logistic map ($D=3$). {\bf{(a)}} shows the whole chaotic region of the logistic map. The portion in green is then zoomed in in {\bf{(b)}}, where the fractal behaviour is evident. {\bf{(c)}} zooms in the rectangle in (b).}
\label{fractal}
\end{figure}

For all the previous figures we used $10^4$ points ($10^4$ different r values). In Fig.~\ref{fractal} we show the logistic map using $10^5$ points in a reduced subset of $r$ values, to see the finer details of the projection. Figure \ref{fractal}a shows the wide-range logistic map projected on the plane. The green section is zoomed in in Fig.~\ref{fractal}b, where the fractal behaviour of the logistic map is noticeable. One can recognize the same pattern repeating as the control parameter $r$ increases and the projection on the plane goes down in FIM, while the pattern on the plane gets smaller and smaller. Figure~\ref{fractal}c zooms in into the rectangular section of Fig.~\ref{fractal}b, to appreciate more detail.

\begin{figure}[ht]
\centerline{
\includegraphics[width=7cm]{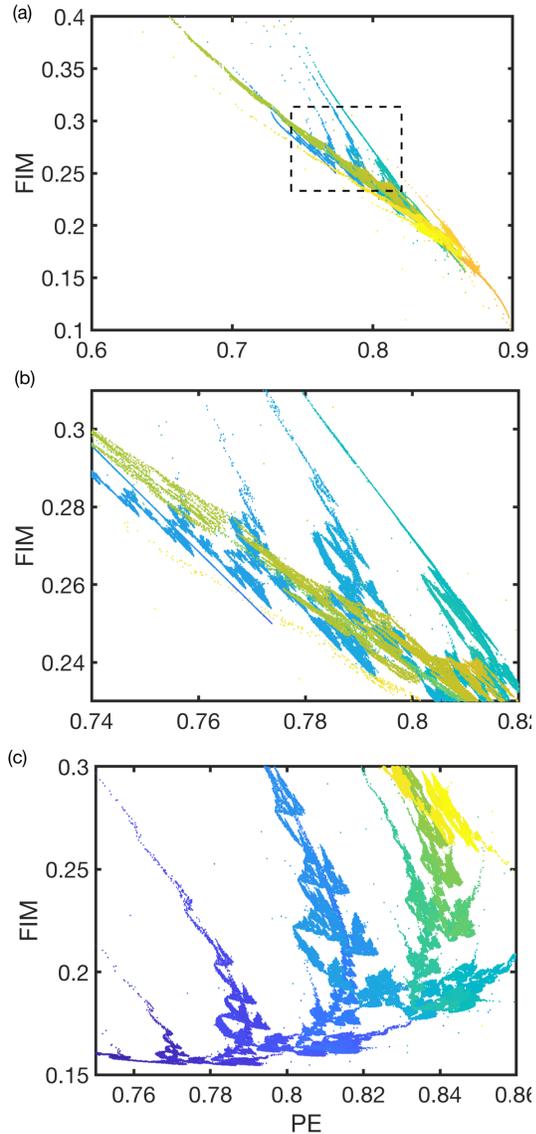}
}
\caption{Fisher-Shannon plane using SA-Lehmer for the logistic map. {\bf{(a)}} Whole chaotic region of the logistic map. The dashed rectangle is zoomed in in {\bf{(b)}}, where the fractal behaviour is clearer. {\bf{(c)}} Fractal behaviour of the logistic map depicted with SA$^{\dagger}$.}
\label{fractal_Lehmer}
\end{figure}

Even though the fractal behaviour of the logistic map, as well as its relation to the Maldelbrot set, is well documented~\cite{1999_PhysicaD_Beck,2001_PRE_Isaeva,2009_CSF_Rani}, it is remarkable that the words can also extract the self-similar essence through the Fisher-Shannon plane. Certainly, to appreciate this, one needs to compute a large density of $r$ values for the projection, but it is worth stressing here that the sorting array can make a difference in this direction, as one can intuitively guess some self-similar behaviour from SA$^*$ (Fig.~\ref{logistic}d) while non can be appreciated from SA-Lehmer (Fig.~\ref{logistic}a).

Once we have seen that, for SA$^*$, the words-Fisher-Shannon plane can detect fractal behaviour in these chaotic maps, we can explore other sorting arrays to look for self-similarity in the dynamics. Figures~\ref{fractal_Lehmer}a and \ref{fractal_Lehmer}b show the complexity plane using SA-Lehmer for $10^5$ points for the logistic map. In it we can also appreciate a repeating pattern in the same range of parameters as in Fig.~\ref{fractal}. The fact that we are using a different sorting array makes the repeating pattern be different in this case, but the underlying dynamical behaviour is clearly captured.

Similar results are found for other sorting arrays for the logistic map, and similar results are found for the other chaotic maps that present fractal behaviour, in this case, through period doubling route to chaos. Figure \ref{fractal_Lehmer}c shows the same fractal behaviour, although with a different pattern on the plane, for the logistic map with SA$^{\dagger}$.

\section{\label{sec:level1}conclusions}

To summarize, we have investigated the relevance of order in the words of dimension $D=3$ and its impact on the Fisher-Shannon plane when computing PE and ordinal-patterns-based FIM. First we have found that interesting features that had been highlighted in recent research is robust to all orders: ordinal-patterns Fisher-Shannon projections help detect and classify different dynamical regions in chaotic maps. Also, the non-invertible maps studied present a common fingerprint on the plane when plotted with the same words order, independently of which order. The landscape of the fingerprint depends on the sorting array but it is the same for all the iterative maps considered. 

Further, whilst higher order analysis ($D=4, 5, 6, \dots $) is able to capture more features of the dynamics, a suitable choice of order for lower dimension ($D=3$) can attain the same level of detail, while saving computational resources.

Looking at the dependence on the control parameter of the words probabilities we have extracted internal symmetries hidden in the chaotic dynamics of the maps. These symmetries are related to time inversion, and rotation symmetries, and impose constraints in the probabilities of the words, even in chaotic regions of the maps. These symmetries allow us to simulate the projection of the maps on the ordinal-patterns Fisher-Shannon plane. The simulations find the allowed region on the plane for the complex dynamics, under those symmetries. Therefore, given a new chaotic system that presents those symmetries we could determine the expected region where to find it on the plane.

By exploring the complexity plane with more informative sorting arrays (such as SA$^*$ and SA$^{\dagger}$) finer details can be appreciated, such as self-similarity. Although fractal behaviour in period-doubling route to chaos systems is well know, it is remarkable how this technique can easily portray it. Also, the fractal behaviour is present independently of the sorting array. Of course, there are some SAs that capture those features in a more clear manner. We want to stress that, while some SAs can present the landscape of the system in a compressed, narrow region, with plenty of overlap, other SAs present the system more spread out, where it is easier to characterize and distinguish different dynamics. This applies also to detecting self-similarity.

Some other interesting aspect of this research that would be of interest to explore is to find a physical or mathematical generic method to determine which SA is optimum, i.e., presents a wider landscape, a broader range for FIM values. We have seen that this helps distinguish and characterize the complex dynamics. We have tried several criteria (for $D=3$), such as minimizing the distance to Gaussian white noise, or maximizing FIM. Although some of them seemed to be better than SA-Lehmer, non of them produced SA$^*$ or SA$^{\dagger}$.

Another interesting aspect to explore is to generate a generalization to $D=4$ of SA$^*$ or SA$^{\dagger}$. For $D=4$ there are 24 words, and more than $6.2\times10^{23}$ different sorting arrays. We have used SA-Lehmer to show that higher dimensions can extract more details of the complex dynamics. It is reasonable to think that there are SAs of $D=4$ that can do a better job.

\section{\label{sec:level1}Data availability}

The data that support the findings of this study are available from the corresponding author upon reasonable request.

\section{\label{sec:level1}Bibliography}


\appendix

\nocite{*}
\bibliography{aipsamp}

@PREAMBLE{
 "\providecommand{\noopsort}[1]{}" 
 # "\providecommand{\singleletter}[1]{#1}%" 
}

@BOOK{2004_Frieden,
   author       = {B. R. Frieden},
   year         = 2004,
   title        = {Science from Fisher information: a unification},
   publisher    = {Cambridge University Press}
}

@ARTICLE{2015_Chaos_Kantz,
   author       = "Elisabeth Bradley and Holger Kantz",
   title        = "Non-linear time seires analysis revisited",
   year         = "2015",
   journal      = "Chaos",
   volume       = "25",
   pages        = "097610",
}

@ARTICLE{2017_PRL_Politi,
   author       = "Antonio Politi",
   title        = "Quantifying the Dynamical Complexity of Chaotic Time Series",
   year         = "2017",
   journal      = "Phys. Rev. Lett.",
   volume       = "118",
   pages        = "144101",
}

@ARTICLE{2020_Nature_Esposito,
   author       = "Daniel Toker and Friedrich T. Sommer and Mark D'Esposito",
   title        = "A simple method for detecting chos in nature",
   year         = "2020",
   journal      = "Nat. Comm. Biology",
   volume       = "3",
   pages        = "",
}

@ARTICLE{2013_SciRep_Aragoneses,
   author       = "Andr\'es Aragoneses and Niclas Rubido and Jordi Tiana-Alsina and M. C. Torrent and Cristina Masoller",
   title        = "Distinguishing signatures of determinism and stochasticity in spiking complex systems",
   year         = "2013",
   journal      = "Sci. Rep.",
   volume       = "3",
   pages        = "1778",
}

@ARTICLE{2003_RSI_Tracy,
   author       = "C. S. Daw and C. E. A. Finney and E. R. Tracy",
   title        = "A review of symbolic analysis of experimental data",
   year         = "2003",
   journal      = "Rev. Sci. Instr.",
   volume       = "74",
   pages        = "915",
}

@ARTICLE{2021_Spichak,
   author       = "David Spichak and Audrey Kupetsky and Andr\'es Aragoneses ",
   title        = "Characterizing complexity of non-invertible chaotic maps in the
Shannon–Fisher information plane with ordinal patterns",
   year         = "2020",
   journal      = "Chaos, solitons, and fractals",
   volume       = "142",
   pages        = "110492",
}

@ARTICLE{2003_Vignat,
   author       = "C. Vignat and J. F. Bercher", 
   title        = "Analysis of signals in the Fisher–Shannon information plane",
   year         = "2003", 
   journal      = "Phys. Lett. A", 
   volume       = "27", 
   pages        = "312",
}

@ARTICLE{2002_PRL_BP,
   author       = "C. Bandt and B. Pompe",
   title        = "Permutation Entropy: A Natural Complexity Measure for Time Series",
   journal      = "Phys. Rev. Lett.",
   volume       = "88", 
   pages        = "174102",
   year         = "2002",
}

@ARTICLE{2012_PhysicaA_Olivares,
   author       = "Felipe Olivares and Angelo Plastino and Osvaldo A. Rosso", 
   title        = "Ambiguities in Bandt–Pompe’s methodology for local entropic quantifiers", 
   journal      = "Physica A", 
   volume       = "391", 
   pages        = "2518--2516", 
   year         = "2012", 
}

@ARTICLE{2014_OptEx_Toomey,
   author       = "J. P. Toomey and D. M. Kane", 
   title        = "Mapping the dynamic complexity of a semiconductor laser with optical feedback using permutation entropy", 
   journal      = "Opt. Express", 
   volume       = "22", 
   pages        = "1713", 
   year         = "2014", 
}

@ARTICLE{2017_Entropy_Bandt,
   author       = "Christoph Bandt", 
   title        = "A New Kind of Permutation Entropy Used to Classify Sleep Stages from Invisible EEG Microstructure", 
   journal      = "Entropy", 
   volume       = "19", 
   pages        = "197", 
   year         = "2017", 
}

@ARTICLE{2016_PRL_Aragoneses,
   author       = "Andr\'es Aragoneses and Laura Carpi and Nikita Tarasov and Dimitri V. Churkin and M. C. Torrent  and Cristina Masoller and Sergei K. Turitsyn", 
   title        = "Unveiling Temporal Correlations Characteristic of a Phase Transition in the Output Intensity of a Fiber Laser", 
   journal      = "Phys. Rev. Lett.", 
   volume       = "116", 
   pages        = "33902", 
   year         = "2016", 
}

@ARTICLE{2018_SciRep_Colet,
   author       = "Meritxell Colet and Andr\'es Aragoneses", 
   title        = "Forecasting events in the complex dynamics of a semiconductor laser with optical feedback", 
   journal      = "Sci. Rep.", 
   volume       = "8", 
   pages        = "10741", 
   year         = "2018", 
}

@ARTICLE{2019_Entropy_Bandt,
   author       = "Christoph Bandt", 
   title        = "Small Order Patterns in Big Time Series: A Practical Guide", 
   journal      = "Entropy", 
   volume       = "21", 
   pages        = "613", 
   year         = "2019", 
}

@ARTICLE{2020_FrontPhys_Zanin,
   author       = "Massimiliano Zanin and Bahar Guntekin and Tuba Akturk and Lutfu Hanoglu and David Papo", 
   title        = "Time Irreversibility of Resting-State Activity in the Healthy Brain and Pathology", 
   journal      = "Front. Physiol.", 
   volume       = "10", 
   pages        = "1619", 
   year         = "2020", 
}

@ARTICLE{1999_PLA_Plastino,
   author       = "M. T. Martin and F. Pennini and A. Plastino", 
   title        = "Analysis of Shannon-Fisher information plane in time series based on information entropy", 
   journal      = "Phys. Lett. A", 
   volume       = "256", 
   pages        = "173 –- 80", 
   year         = "1999", 
}

@ARTICLE{2012_Phys_LettA_Olivares,
   author       = "Felipe Olivares and Angelo Plastino and Osvaldo A. Rosso", 
   title        = "Analysis of Shannon-Fisher information plane in time series based on information entropy", 
   journal      = "Phys. Lett. A", 
   volume       = "376", 
   pages        = "1577 -- 1583", 
   year         = "2012", 
}

@ARTICLE{2014_PLOS_Carpi,
   author       = "Mart\'in G\'omez Ravetti and Laura C. Carpi and Bruna Amin Goncalves and Alejandrop C. Freri and Osvaldo A. Rosso", 
   title        = "Distinguishing Noise from Chaos: Objective versus Subjective Criteria Using Horizontal Visibility Graph", 
   journal      = "PLOS ONE", 
   volume       = "9", 
   pages        = "108004", 
   year         = "2014", 
}

@ARTICLE{2018_PRA_Carpi,
   author       = "Laura Carpi and Cristina Masoller", 
   title        = "Persistence and stochastic periodicity in the intensity dynamics of a fiber laser during the transition to optical turbulence", 
   journal      = "Phys. Rev. A", 
   volume       = "97", 
   pages        = "023842", 
   year         = "2018", 
}

@ARTICLE{1925_Fisher,
   author       = "R. A. Fisher", 
   title        = "Theory of statistical estimation", 
   journal      = "Proc. Camb. Phil. Soc.", 
   volume       = "22", 
   pages        = "700 –- 25", 
   year         = "1925", 
}

@MISC{logistic_video,
   title        = "https://bit.ly/2XEqqvm",
year         = "2021", 
}

@MISC{logistic_cusp_video,
   title        = "https://www.youtube.com/watch?v=xntTd3P_-LA",
year         = "2021", 
}

@ARTICLE{1999_PhysicaD_Beck,
   author       = "Christian Beck", 
   title        = "Physical meaning for Maldelbrot and Julia sets", 
   journal      = "Physica D", 
   volume       = "125", 
   pages        = "171 –- 182", 
   year         = "1999", 
}

@ARTICLE{2001_PRE_Isaeva,
   author       = "Olga B. Isaeva and Sergei P. Kuznetsov and Vladimir I. Ponomarenko", 
   title        = "Mandelbrot set in coupled logistic maps and in an electronic experiment ", 
   journal      = "Phys. Rev. E", 
   volume       = "64", 
   pages        = "055201", 
   year         = "2001", 
}

@ARTICLE{2009_CSF_Rani,
   author       = "Mamta Rani and Rashi Agarwal", 
   title        = "Generation of fractals from complex logistic map", 
   journal      = "Chaos, Solitons, and Fractals", 
   volume       = "42", 
   pages        = "447 -- 452", 
   year         = "2009", 
}

@ARTICLE{2021_Gunther,
   author       = "Ivan Gunther and Arjendu K. Pattanayak and Andr\'es Aragoneses", 
   title        = "Ordinal patterns in the Duffing oscillator: Analyzing powers of characterization", 
   journal      = "Chaos", 
   volume       = "31", 
   pages        = "023104", 
   year         = "2021", 
}

@CONTROL{REVTEX41Control}

@CONTROL{aip41Control,pages="1",title="0"}

\end{document}